
======================================================================

\magnification=\magstep1
\baselineskip=24pt
\hsize = 6 true in
\pretolerance=10000
\def\n{\noindent}

\def\b{\bigskip}
\def\c{\centerline}
\def\n{\noindent}
\def\s{\smallskip}

\line{\hfil IUCAA- 2 /93 Feb'93}

\vskip 2 cm

\c{\bf Singularity Free Inhomogeneous Models with Heat Flow}
\vskip 1.5 cm
\c{\bf L. K. Patel$^*$ and Naresh Dadhich$^{**\dag}$}

\c{$^*$Department of Mathematics, Gujarat University, }
\c{ Ahmedabad - 380 009, India}

\vskip 1cm

\c{$^{**}$ Inter University Centre for Astronomy and Astrophysics,}
\c{ Post Bag 4, Ganeshkhind, Pune - 411 007.}

\vskip 1 cm

\c{\bf Abstract}

\vskip 0.5 cm

\n We present a class of singularity free exact cosmological
solutions of Einstein's equations describing a perfect
fluid with heat flow. It is obtained as generalization of
the Senovilla class [1] corresponding to incoherent
radiation field. The spacetime is cylindrically symmetric and globally
regular.

\b

\n PACS numbers : 04.20 Jb, 98.80 Dr
\b
\n $\dag$ E-mail address : naresh@iucaa.ernet.in
\vfill\eject

\n Senovilla [1] has opened up a new vista of singularity free cosmological
models. He obtained a class of cyclindrically symmetric inhomogeneous
exact solutions of Einstein's equations describing a perfect fluid
with the equation of state $\rho = 3p$. The spacetime was
globally regular, smooth and satisfied the energy and causality conditions.
All the physical and curvature invariants were finite everywhere,
and all causal curves were complete [2]. There was no singularity of any kind.

\s

\n The singularity structure of a general class of inhomogeneous perfect
fluid solutions was analysed [3] and it was shown that existence or
non-existence of singularity as well as its kind (spacelike,
timelike, Weyl, Ricci, etc) depended upon the particular choice
of certain parameters and metric functions. There exists a general class of
inhomogeneous perfect fluid solutions without singularity. Recently
we have also found [4] a new class of solutions for stiff fluid
$(\rho = p)$ that have vacuum spacetime as the matter free $(\rho = 0)$
limit. It is very curious that spacetime is empty as well as without
singularity. One
wonders about the source of its curvature.

\s

\n How robust is the singularity free character of spacetime ? In particular
what happens when dissipative effects of viscosity, heat flow etc. are
incorporated ?
In this note we wish
to obtain a generalization  of the Senovilla [1] model that includes heat flow.

\s

\n Dissipative effects in cosmology were considered in the context of large
entropy per baryon and isotropy of microwave background radiation[5,6] and heat
flow
was in particular investigated by several authors [7 - 16]. In very
early times, matter is supposed to be highly dense and hot, and hence
consideration of heat flow for very early evaluation of the Universe
is quite appropriate.

\s

\n The energy momentum tensor incorporating heat flow with perfect fluid reads
as

$$T_{ik} =(\rho + p) u_i u_k - pg_{ik} + 2 u_{(i} q_{k)} \eqno(1) $$

\n where $u_iu^i = 1, q_i u^i = 0, $  and
$q_i$ is the heat flow vector. The expansion $\theta$, shear scalar $\sigma$
and the
acceleration vector $\dot u_i$ for the fluid are defined by

$$\eqalign{\theta  &= u^i_{; i} ,\quad \sigma^2 = \sigma_{ik} \sigma^{ik}\cr
\sigma_{ik} &= u_{(i;k)} - u_{(i}\dot u_{k)} - {\theta \over 3}(g_{ik} -
u_iu_k) \cr
\dot u_i &= u_{i;k} u^k}. $$


\n For Einstein's equations

$$R_{ik} - {1 \over 2} R g_{ik} = -8 \pi T_{ik} \eqno(2) $$

\n we obtain the following class of solutions for perfect fluid
with radial heat flow;

$$ ds^2 = A^2 (dt^2 - dr^2) - B^2dz^2 - D^2 d \phi^2 \eqno(3) $$

\n where

$$\eqalign{A &=C^{1-b}(at)  \quad C^{(b+4)/(1-2b)}(3ar)\cr
B &=C^b(at) \quad C^{-(b+4)/9}(3ar)\cr
D &={1\over 3a} S(3ar)\quad C^{-(b+4)/9}(3ar)\quad C^{(1-b)}(at). \cr}
\eqno(4)$$


\n Here $a $ and $b$ are free parameters and we denote $C(x) = coshx, S(x) =
sinhx,
 T(x) = tanhx,
Se(x) = sechx$. The metric (3) is globally regular for
entire range of the coordinates; $- \infty < t, z < \infty, 0 < r < \infty,
0 < \phi < 2 \pi$. The spacetime is cylindrically symmetric without
any singularity.
\s

\n The physical and kinematical parameters have the following explicit
expressions:

$$8 \pi \rho A^2 = \left( {a^2 \over 3(1-2b)} \right) (b+4) (2b + 17)
(2 - b) Se^2(3ar)$$
$$+ {a^2 \over 3} (5 - b) (b+1) + a^2 (1 - b^2) T^2(at)
\eqno(5.1) $$

$$ 8 \pi p A^2 = \left({ a^2 \over 9(1 - 2b)}\right) ( b + 4)^2 (2b +17)
Se^2(3ar) + {a^2 \over 9} (b+1) (b+7)$$
$$ + a^2 (1 - b^2) T^2(at) \eqno(5.2)$$

$$\eqalignno{ 8 \pi q_{r} A &= {4a^2 \over 3(1-2b)} (b+4) (2 - b) (b +1) T(at)
T(3ar) &(5.3)\cr
\theta A &= a (2 -b) T(at) &(5.4)\cr
\sigma^2A^2 &= {2 \over 3} a^2(1 - 2b)^2 T^2(at) &(5.5) \cr
\dot u_{ r} &= - {3a \over 1-2b} (b+4) T(3ar). &(5.6)\cr} $$

\n The phenomenological expression for the heat conduction is given by

$$q_{_k} = \psi (F,_{i} + F \dot u_{i}) (\delta^i_k - u^i u_k ) \eqno(6) $$

\n where  $\psi $ is the thermal conductivity and $F$ is the temperture.
 For the case under consideration, the above equation can be integrated if
$\psi$ is
a function of $t$ alone. The temperature distribution in the Universe is then
given by

$$F = l_0(t) C^{(b+4)/(1 - 2b)}(3ar) - {  a \over 36\pi \psi} (2-b)(b+1)
T(at)~~C^{-(b + 4)/(1-2b)}(3ar) \eqno(7) $$

\n where $l_0$ is an arbitrary function of $t$.
\s
\n The solution(3) reduces to the Senovilla [1] perfect fluid model for
$b = -1$, when heat flux is switched off $(q_{_k} = \psi = 0)$. From the
expressions in equation (5) the physical requirements $(\rho >  p >0, ~~ \rho +
3p > 0 )$ restrict the range for $b$ to
$-1 \leq b < 1/2$. All the parameters remain finite and well behaved
for all values of $t$ and $r$. We now compute the
Weyl scalars and they are as follows~:

$$\eqalign{\psi_0 + \psi_4 =&C^{2(b-1)}_{~~~~(at)} C^{{2(b+4)\over
2b-1}}_{~~~~~(3ar)}
\bigg[{a^2 \over (1-2b)} (39 + 5b - 4b^2)\cr
&+ (1-b) (1-2b) a^2 T^2 (at)\bigg]\cr}$$

$$\psi_0 - \psi_4 = {2\over 3} a^2 (5-b) (b + 4) T(at) T(3ar)
C^{2(b-1)}_{~~~~(at)} C^{{2(b+4)\over 2b-1}}_{~~~~~(3ar)}$$

$$\eqalign{\psi_2 =&{1\over 6} C^{2(b-1)}_{~~~~(at)}
C^{{2(b+4)\over 2b-1}}_{~~~~~(3ar)}
\bigg[ {a^2 \over 1 - 2b} (35 + 13b - 4b^2)\cr
&+(1 - 3b + 2b^2) a^2 T^2(at) - {2 (20 + b - b^2 \over (1 - 2b)} a^2 T^2
(3ar)\bigg]\cr}$$

It is clear from the above expressions that the Weyl scalars remain
regular and finite for the entire range of coordinates for $-1 \leq b < {1\over
2}$.

\s

\n From relations (5) it follows that as $t \rightarrow - \infty$
all the physical and
kinematic parameters, as well as the Weyl curvatures
tend to zero, however the metric does not go over to flat spacetime
(this also happens in [1]). Then as $t$ increases, the fluid
starts contracting , the density and pressure increase reaching the maximum
at $t = 0$. The heat flux $q_{r}$ and expansion $\theta$ are negative
for $- \infty < t < 0$, zero at $t = 0$ and positive for $0 < t < \infty$.
Similarly the shear is positive in $r$ and $\phi$ directions and negative in
$z$ direction and it reverses its sense at $t = 0$. The radial acceleration is
zero
at $r = 0$ and negative for $r > 0$. As $t$ increases from zero, the fluid
starts expanding and heat flows radially out. All physical and kinematical
parameters tend to zero  again as $t \rightarrow \infty$ and the spacetime
tends to
the same state as that of $t \rightarrow \infty$.
The universe therefore starts from nearly flat spacetime at $t \rightarrow -
\infty,$
passes through the dense and hot state at $t = 0$ (the free parameter $a$ can
be
chosen as large as we please) and becoming nearly
flat and cold  again as $t \rightarrow \infty$.

We have thus shown that it is possible to incorporate heat flux in the
Senovilla class [1] of models without disturbing their singularity free
character.
We have also subsequently generalised the general class of perfect fluid
solution
[3] to include heat flux [17].

\vfill\eject

\n{\bf References}
\b

\item{[1]} Senovilla J.M.M., 1990, Phys. Rev. Lett. {\it 64}, 2219.
\s
\item{[2]} Chinea, F.J., Fernandez-Jambrina, L. and Senovilla, J.M.M., 1992,
Phys. Rev., {\it D45}, 481.
\s
\item{[3]}Ruiz, E. and Senovilla, J.M.M., 1992, Phys. Rev. {\it D45}, 1995.
\s
\item{[4]} Patel L.K. and Dadhich N., 1992 ,Preprint IUCAA-1/93 Jan'93

\s
\item{[5]} Misner, C.W., 1968, Ap. J. {\it 151}, 431.
\s
\item{[6]} Caderni, N. and Fabri, R., 1978, Nuovo Cimento, {\it 44B}, 228.
\s
\item{[7]}Dang, Y., 1989, Gen. Rel. Grav., {\it 21}, 503.
\s
\item{[8]}Mukherjee, G., 1986, J. Astrophys. Astron., {\it 7}, 259.
\s
\item{[9]}Novello, M. and Reboucas, M.J., 1978, Ap. J., {\it 225}, 719.
\s
\item{[10]} Ray, D., 1980, J. Math. Phys., {\it 21}, 2797.
\s
\item{[11]} Reboucas, M.J. and de Limma, J.A.S., 1982, J. Math. Phys. {\it 23},
1547.
\s
\item{[12]} Reboucas, M.J. and de Limma, J.A.S.,  1981, J. Math. Phys. {\it
22},
2699.
\s
\item{[13]}Reboucas, M.J., 1980, Nuovo Cimento, {\it 67B}, 120.
\s
\item{[14]} Bradley, J.M. and Sveistins, E., 1984, Gen. Rel. Grav. {\it 16},
1119.
\s
\item{[15]} Sveistins, E., 1985, Gen. Rel. Grav., {\it 17}, 521.
\s
\item{[16]}Patel, L.K. and Dadhich, N., 1991, Ap.J. {\it 401}, 433.
\s
\item{[17]}Patel, L.K. and Dadhich, N., to be published.

\bye